\documentclass{llncs}

\usepackage{graphicx}
\usepackage{caption}
\usepackage{subcaption}
\usepackage{mathtools}
\usepackage{multirow}
\usepackage{algorithm}
\usepackage[noend]{algpseudocode}
\usepackage[utf8x]{inputenc}
\usepackage[dvipsnames]{xcolor}
\usepackage{pgfplots}
\pgfplotsset{compat=1.14}
\usepackage[hidelinks=true, pdffitwindow=true, setpagesize=false]{hyperref}
\usepackage{listings}
\usepackage{color}

\lstdefinestyle{mystyle}{
	backgroundcolor=\color{white},   
	commentstyle=\color{green},
	keywordstyle=\color{orange},
	numberstyle=\tiny\color{black},
	stringstyle=\color{red},
    basicstyle=\small,
	breakatwhitespace=false,         
	breaklines=true,                 
	captionpos=b,                    
	keepspaces=true,                 
	numbers=none,
	numbersep=5pt,                  
	showspaces=false,                
	showstringspaces=false,
	showtabs=false,                  
	tabsize=2,
	frame=none    
}

\usepackage{ifsym}

\begin{document}
	
	\title{T$^2$K$^2$: The Twitter Top-K Keywords Benchmark}
	
	\author{Ciprian-Octavian Truică$^{1, a}$, J{\'e}r{\^o}me Darmont$^{2, b}$}
	
	\institute{
		$^{1}$Computer Science and Engineering Department, Faculty of Automatic Control and Computers, University Politehnica of Bucharest, Bucharest, Romania\\
		$^{2}$Universit\'{e} de Lyon, Lyon 2, ERIC EA 3083, France\\
		$^{a}$ciprian.truica@cs.pub.ro, $^{b}$jerome.darmont@univ-lyon2.fr 
	}
	\maketitle
	
	\begin{abstract}
   	Information retrieval from textual data focuses on the construction of vocabularies that contain weighted term tuples. Such vocabularies can then be exploited by various text analysis algorithms to extract new knowledge, e.g., top-k keywords, top-k documents, etc. Top-k keywords are casually used for various purposes, are often computed on-the-fly, and thus must be efficiently computed. To compare competing weighting schemes and database implementations, benchmarking is customary. To the best of our knowledge, no benchmark currently addresses these problems. Hence, in this paper, we present a top-k keywords benchmark, T$^2$K$^2$, which features a real tweet dataset and queries with various complexities and selectivities. T$^2$K$^2$ helps evaluate weighting schemes and database implementations in terms of computing performance. To illustrate T$^2$K$^2$'s relevance and genericity, we show how to implement the TF-IDF and Okapi BM25 weighting schemes, on one hand, and relational and document-oriented database instantiations, on the other hand.
	\end{abstract}
	
	\keywords{Top-k keywords, Benchmark, Term weighting, Database systems}
	
	\section{Introduction}    
    
    Analyzing textual data is a current challenge, notably due to the vast amount of text generated daily by social media. One approach for extracting knowledge is to infer from texts the top-k keywords to determine trends \cite{Bringay2011,Ravat2008}, or to detect anomalies or more generally events \cite{Guille2015}. Computing top-k keywords requires building a weighted vocabulary, which can also be used for many other purposes such as topic modeling and clustering.
    Term weights can be computed at the application level, which is inefficient when working with large data volumes because all information must be queried and processed at a layer different from storage. A presumably better approach is to process information at the storage layer using aggregation functions, and then return the result to the application layer. Yet, the term weighting process remains very costly, because each time a query is issued, at least one pass through all documents is needed. 
    
    To compare combinations of weighting schemes, computing strategies and physical implementations, benchmarking is customary. However, to the best of our knowledge, there exists no benchmark for this purpose. Hence, we propose in this paper the Twitter Top-K Keywords Benchmark (T$^2$K$^2$), which features a real tweet dataset and queries with various complexities and selectivities. We designed T$^2$K$^2$ to be somewhat generic, i.e., it can compare various weighting schemes, database logical and physical implementations and even text analytics platforms \cite{Truica2016} in terms of computing efficiency. As a proof of concept of T$^2$K$^2$'s relevance and genericity, we show how to implement the TF-IDF and Okapi BM25 weighting schemes, on one hand, and relational and document-oriented database instantiations, on the other hand.
               
    The remainder of this paper is organized as follows. Section~\ref{sec:ralted_work} reviews text-oriented benchmarks. Section~\ref{sec:approach} provides T$^2$K$^2$'s generic specification.  Section~\ref{sec:validation} details T$^2$K$^2$'s proof of concept, i.e., its instantiation for several weighting schemes and database implementations. Finally, Section~\ref{sec:conclusion} concludes this paper and hints at future research. 
    
    \section{Related Works}~\label{sec:ralted_work}

    Term weighting schemes are extensively benchmarked in sentiment analysis \cite{Reagan2015}, semantic similarity \cite{OShea2010}, text classification and categorization \cite{Kilinc2017,Lewis2004,OShea2010,Partalas2015}, and textual corpus generation \cite{Wang2016}. Benchmarks for text analysis focus mainly on algorithm accuracy, while either term weights are known before the algorithm is applied, or their computation is incorporated with preprocessing. Thus, such benchmarks do not evaluate weighting scheme construction efficiency as we do.   
    
    Other benchmarks evaluate parallel text processing in big data applications in the cloud  \cite{Ferrarons2014,Gattiker2013}. PRIMEBALL notably specifies several relevant properties characterizing cloud platforms \cite{Ferrarons2014}, such as scale-up, elastic speedup, horizontal scalability, latency, durability, consistency and version handling,  availability, concurrency and other data and information retrieval properties. However, PRIMEBALL is only a specification; it is not implemented.    
    
    \section{T$^2$K$^2$ Specification}~\label{sec:approach}
   
	Typically, a benchmark  is  constituted  of a  data  model  (conceptual  schema  and extension), a  workload  model (set  of operations) to  apply  on  the dataset, an execution protocol and performance  metrics~\cite{Darmont2014}. In this section, we provide a conceptual description of T$^2$K$^2$, so that it is generic and can cope with various weighting schemes and database logical and physical implementations.
       
    \subsection{Data Model}

    The base dataset we use is a corpus of 2\,500\,000 tweets that was collected using Twitter's REST API to read and gather data. Moreover, we applied preprocessing steps to the raw corpus to extract the additional information needed to build a weighted vocabulary: 
1) extract all tags and remove links;
2) expand contractions, i.e., shortened versions of the written and spoken forms of a word, syllable, or word group, created by omission of internal letters and sounds~\cite{Cooper2014}, e.g., "it's" becomes "it is"; 
3) extract sentences and remove punctuation in each sentence, creating a clean text;
4) for each sentence, extract lemmas and create a lemma text;
5) for each lemma $t$ in tweet $d$, compute the number of co-occurrences $f_{t,d}$ and
term frequency $TF(t,d)$, which normalizes $f_{t,d}$.

\begin{figure}[!hbtp]
		\begin{center}
		\includegraphics[width=\columnwidth]{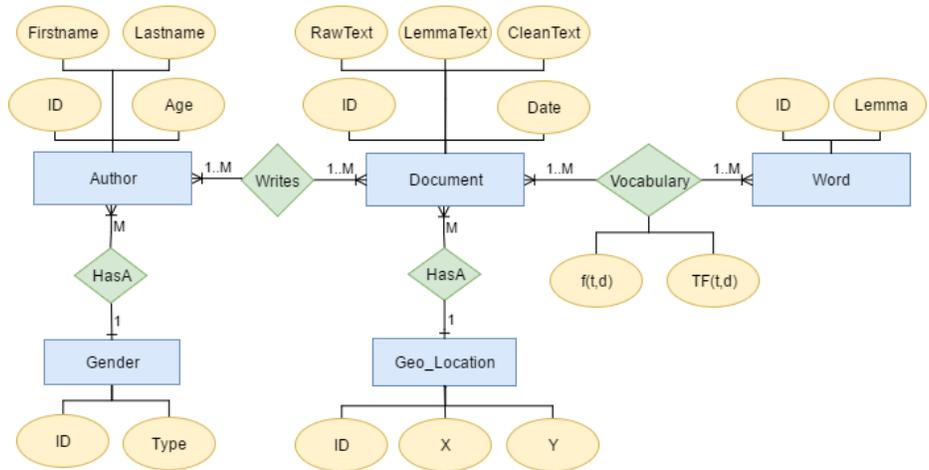}
		\caption{T$^2$K$^2$ Conceptual Data Model} 
		\label{fig:erdiagram}
		\end{center}
	\end{figure}

 T$^2$K$^2$ database's conceptual model (Figure~\ref{fig:erdiagram}) represents all the information extracted after the text preprocessing steps. 
Information about tweet \textit{Author} are a unique identifier, first name, last name and age.
Information about author \textit{Gender} is stored in a different entity to minimize the number of duplicates of gender type.
\textit{Documents} are identified by the tweet's unique identifier and store the raw tweet text, clean text, lemma text, and the tweet's creation date.
\emph{Writes} is the relationship that associates a tweet to its author.
Tweet location is stored in the \emph{Geo\_Location} entity to avoid duplicates again. 
\textit{Word} bears a unique identifier and the actual lemma.
Finally, weights $f_{t,d}$ and $TF(t,d)$ for each lemma and each document are stored in the \emph{Vocabulary} relationship.

    The initial 2\,500\,000 tweet corpus is split into 5 different datasets that all keep an equal balance between the number of tweets for both genders, location and date. These datasets contain 500\,000, 1\,000\,000, 1\,500\,000, 2\,000\,000 and 2\,500\,000 tweets, respectively. They allow scaling experiments and are associated to a scale factor ($SF$) parameter, where $SF \in \{0.5, 1, 1.5, 2, 2.5\}$, for conciseness sake.     
    
    \subsection{Workload Model}
    \label{sec:workload}
    
    The queries used in T$^2$K$^2$ are designed to achieve two goals:
1) compute different term weighting schemes using aggregation functions and return the top-k keywords;
2) test the performance of different database management systems.
T$^2$K$^2$ queries are sufficient for achieving these goals, because they test the query execution plan, internal caching and the way they deal with aggregation. More precisely, they take  different group by attributes into account and aggregate the information to compute weighting schemes for top-k keywords.
 
	T$^2$K$^2$ features four queries $Q1$ to $Q4$ that compute top-k keywords w.r.t. constraint(s): $c_1 (Q1)$, $c_1 \wedge c_2 (Q2)$, $c_1 \wedge c_3 (Q3)$, $c_1 \wedge c_2 \wedge c_3 (Q4)$. 
$c_1$ is \emph{Gender.Type = pGender}, where parameter  \emph{pGender $\in$ \{male, female\}}.
$c_2$ is  \emph{Document.Date $\in$ [pStartDate, pEndDate]}, where \emph{pStartDate, pEndDate $\in$ [2015-09-17 20:41:35, 2015-09-19 04:05:45]} and \emph{pStartDate $<$ pEndDate}.
$c_3$ is \emph{Geo\_location.X $\in$ [ pStartX, pEndX]} and \emph{Geo\_location.Y $\in$ [pStartY, pEndY]}, where \emph{pStartX, pEndX $\in$ [15, 50]}, \emph{pStartX $<$ pEndX}, \emph{pStartY, pEndY $\in$ [-124, 120]} and \emph{pStartY $<$ pEndY}.
Queries bear different levels of complexity and selectivity. 
        
\subsection{Performance Metrics and Execution Protocol}     
    
	We use  each query's response time $t(Q_i)$ as metrics in T$^2$K$^2$. Given scale factor $SF$, all queries $Q1$ to $Q4$ are executed 40 times, which is sufficient according to the central limit theorem. Average response times and standard deviations are computed for $t(Q_i)$. All executions are warm runs, i.e., either caching mechanisms must be deactivated, or a cold run of $Q1$ to $Q4$ must be executed once (but not taken into account in the benchmark's results) to fill in the cache. Queries must be written in the native scripting language of the target database system and executed directly inside said system using the command line interpreter.     
               
    \section{T$^2$K$^2$ Proof of Concept}~\label{sec:validation}
    
	In this section, we aim at illustrating how T$^2$K$^2$ works and at demonstrating that it can adequately benchmark what it is designed for, i.e., weighting schemes and database implementations. For this sake, we first compare the TF-IDF and Okapi BM25 weighting schemes in terms of computing efficiency. Second, we seek to determine whether a document-oriented database is a better solution than in a relational databases when computing a given term weighting scheme.
        
    \subsection{Weighting Schemes}
    \label{sec:weightingSchemes}
    
    Let $D$ be the corpus of tweets, $N=|D|$ the total number of documents (tweets) in $D$ and  $n$ the number of documents where some term $t$ appears. The TF-IDF weight is computed by multiplying the augmented term frequency $TF(t,d) = K + (1 - K) \cdot \frac{f_{t,d}}{\max_{t' \in d}(f_{t',d})}$) by the inverted document frequency $IDF(t,D) = 1 + \log\frac{N}{n}$, i.e., $TFIDF(t,d,D) = TF(t,d) \cdot IDF(t,D)$. The augmented form of $TF$ prevents a bias towards long tweets when the free parameter $K$ is set to $0.5$~\cite{Paltoglou2010}. It uses the number of co-occurrences $f_{t,d}$ of a word in a document,  normalized with the frequency of the most frequent term $t'$, i.e., $\max_{t' \in d}(f_{t',d})$.
  
 	 The Okapi BM25 weight is given in Equation~\eqref{eq:okapi}, where $||d||$ is $d$'s length, i.e., the number of terms appearing in $d$. Average document length $avg_{d' \in D}(||d'||)$ is used to remove any bias towards long documents. The values of free parameters $k_1$ and $b$ are usually chosen, in absence of advanced optimization, as $k_1 \in [1.2,2.0]$ and $b=0.75$  \cite{Manning2008,SparckJones2000a,SparckJones2000b}.

  \begin{equation}\label{eq:okapi}
      Okapi(t,d,D) = \frac{TFIDF(t,d,D) \cdot (k_1 + 1)}{TF(t,d) + k_1 \cdot (1 - b + b \cdot \frac{||d||}{ avg_{d' \in D}(||d'||)})}
  \end{equation}

    The sum $S\_TFIDF(t,d,D) = \sum_{i=1}^{N} TFIDF(t,d_i,D)$ of all TF-IDFs and the sum $S\_Okapi(t,d,D) = \sum_{i=1}^{N} Okapi(t,d_i,D)$ of all Okapi BM25 weights  constitute the term's weights that are used to construct the list of top-k keywords.    
    
    \subsection{Relational Implementations}
    
    \subsubsection{Database} 
    
\begin{figure}[!hbtp]
		\begin{center}
		\includegraphics[width=\columnwidth]{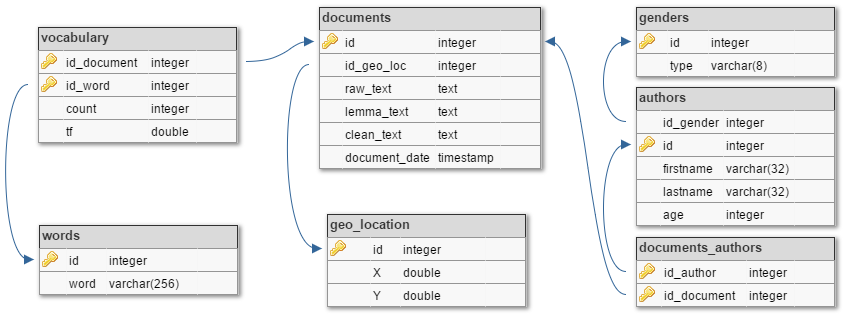}
		\caption{T$^2$K$^2$ Relational Logical Schema} 
		\label{fig:dbscheam}
		\end{center}
	\end{figure}

    The logical relational schema used in both relational databases management systems (Figure~\ref{fig:dbscheam}) directly translates the conceptual schema from Figure~\ref{fig:erdiagram}. 
    
    \subsubsection{Queries} 
Text analysis deals with discovering hidden patterns from texts. In most cases, it is useful to determine such patterns for given groups, e.g., males and females, because they have different interests and talk about disjunct subjects. Moreover, if new events appear, depending on the location and time of day, these subject can change for the same group of people. The queries we propose aim to determine such hidden patterns and improve text analysis and anomaly detection.

    Let us express T$^2$K$^2$'s queries  in relational algebra. $c_1$, $c_2$ and $c_3$ are the constraints defined in Section~\ref{sec:workload}, adapted to the relational schema. 
        
        \emph{Q1 = $\gamma_L$( $\pi_{documents.id, words.word, f_w(vocabulary.count, vocabulary.tf)}$( $\sigma_{c_1}$( documents $\bowtie_{c_4}$ documents\_authors $\bowtie_{c_5}$ authors $\bowtie_{c_6}$ genders $\bowtie_{c_7}$ vocabulary $\bowtie_{c_8}$ words)))}, where $c_4$ to $c_8$ are join conditions; $f_w$ is the weighting function that computes TF-IDF or Okapi BM25, which takes two  parameters: $vocabulary.count$ $=f_{t,d}$ and $vocabulary.tf = TF(t,d)$; $\gamma_L$ is the aggregation operator, where $L=(F, G)$, with $F=$ sum($f_w(vocabulary.count, vocabulary.tf)$) and $G$ is the $words.word$ attribute that appears in the group by clause.
    	
        \emph{Q2 = $\gamma_L$( $\pi_{documents.id, words.word, f_w(vocabulary.count, vocabulary.tf)}$( $\sigma_{c_1 \wedge c_2}$( documents $\bowtie_{c_4}$ documents\_authors $\bowtie_{c_5}$ authors $\bowtie_{c_6}$ genders $\bowtie_{c_7}$ vocabulary $\bowtie_{c_8}$ words)))}.
    	
        \emph{Q3 = $\gamma_L$( $\pi_{documents.id, words.word, f_w(vocabulary.count, vocabulary.tf)}$( $\sigma_{c_1 \wedge c_3}$( documents $\bowtie_{c_4}$ documents\_authors $\bowtie_{c_5}$ authors $\bowtie_{c_6}$ genders $\bowtie_{c_7}$ vocabulary $\bowtie_{c_8}$ words $\bowtie_{c_9}$ geo\_location)))}, where $c_9$ is the join condition between  \emph{documents} and \emph{geo\_location}.
    
    \emph{Q4 = $\gamma_L$( $\pi_{documents.id, words.word, f_w(vocabulary.count, vocabulary.tf)}$( $\sigma_{c_1 \wedge c_2 \wedge c_3}$( documents $\bowtie_{c_4}$ documents\_authors $\bowtie_{c_5}$ authors $\bowtie_{c_6}$ genders $\bowtie_{c_7}$ vocabulary $\bowtie_{c_8}$ words $\bowtie_{c_9}$ geo\_location)))}.

    \subsection{Document-oriented Implementation}
    
    \subsubsection{Database}
    
   In a Document Oriented Database Management System (DODBMS), all information is typically stored in a single collection. The many-to-many \emph{Vocabulary} relationship from Figure~\ref{fig:erdiagram} is modeled as a nested document for each record. The information about user and date become single fields in a document, while the location becomes an array. Figure~\ref{fig:docexample} presents an example of the DODBMS document.

\begin{figure}[!hbtp]
\centering      
\begin{tabular}{|c|}
\hline
\begin{lstlisting}[language=Java]
{   _id : 644626677310603264, 
    rawText : "Amanda's car is too much for my headache", 
    cleanText : "Amanda is car is too much for my headache", 
    lemmaText : "amanda car headache", 
    author : 970993142, 
    geoLocation : [ 32, 79 ], 
    gender : "male", 
    age : 23,
    lemmaTextLength : 3,
    words : [ { "tf" : 1, "count" : 1, "word" : "amanda"}, 
            { "tf" : 1, "count" : 1, "word" : "car" }, 
            { "tf" : 1, "count" : 1, "word" : "headache"} ], 
    date : ISODate("2015-09-17T23:39:11Z") }
\end{lstlisting} \\ \hline
\end{tabular}
\caption{Sample DODBMS Document}
\label{fig:docexample}
\end{figure}
   
\subsubsection{Queries}\label{sec:mongoqs}
    
   In DODBMSs, user-defined (e.g., JavaScript) functions are used to compute top-k keywords. The TF-IDF weight can take advantage of both native database aggregation (NA) and MapReduce (MR). However, due to the multitude of parameters involved and the calculations needed for the Okapi BM25 weighting scheme, the NA method is usually difficult to develop. Thus, we recommend to only use  MR in benchmark runs.
   
\section{Conclusion}~\label{sec:conclusion}

Jim Gray defined four primary criteria to specify a "good" benchmark \cite{Gray1993}.
\textit{Relevance:} The benchmark must deal with aspects of performance that appeal to the largest number of users. Considering the wide usage of top-k queries in various text analytics tasks, we think T$^2$K$^2$ fulfills this criterion. We also show in Section~\ref{sec:validation} that our benchmark achieves what it is designed for.

\textit{Portability:} The benchmark must be reusable to test the performances of different database systems. We successfully instantiated  T$^2$K$^2$ within two types of database systems, namely relational and document-oriented systems.

\textit{Simplicity:} The benchmark must be feasible and must not require too many resources. We designed T$^2$K$^2$ with this criterion in mind (Section~\ref{sec:approach}), which is particularly important for reproducibility. We notably made up parameters that are easy to setup.

\textit{Scalability:} The benchmark must adapt to small or large computer architectures. By introducing scale factor $SF$, we allow users to simply parameterize T$^2$K$^2$ and achieve some scaling, though it could be pushed further in terms of data volume. 

		In future work, we plan to expand T$^2$K$^2$'s dataset significantly to aim at big data-scale volume. We also intend to further our proof of concept and validation efforts by benchmarking other NoSQL database systems and gain insight regarding their capabilities and shortcomings. We also plan to adapt T$^2$K$^2$ so that it runs in the Hadoop and Spark environments.

   	\bibliographystyle{splncs03}
	\bibliography{lncs}

\begin{thebibliography}{10}
\providecommand{\url}[1]{\texttt{#1}}
\providecommand{\urlprefix}{URL }

\bibitem{Bringay2011}
Bringay, S., B{\'e}chet, N., Bouillot, F., Poncelet, P., Roche, M., Teisseire,
  M.: Towards an on-line analysis of tweets processing. In: International
  Conference on Database and Expert Systems Applications (DEXA). pp. 154--161
  (2011)

\bibitem{Cooper2014}
Cooper, J.D., Robinson, M.D., Slansky, J.A., Kiger, N.D.: Literacy: Helping
  students construct meaning. Cengage Learning (2014)

\bibitem{Darmont2014}
Darmont, J.: Data Processing Benchmarks, pp. 146--152. Encyclopedia of
  Information Science and Technology (3rd Edition), IGI Global, Hershey, PA,
  USA (2014)

\bibitem{Ferrarons2014}
Ferrarons, J., Adhana, M., Colmenares, C., Pietrowska, S., Bentayeb, F.,
  Darmont, J.: {PRIMEBALL:} a parallel processing framework benchmark for big
  data applications in the cloud. In: 5th TPC Technology Conference on
  Performance Evaluation and Benchmarking (TPCTC 2013). LNCS, vol. 8391, pp.
  109--124 (2014)

\bibitem{Gattiker2013}
Gattiker, A.E., Gebara, F.H., Hofstee, H.P., Hayes, J.D., Hylick, A.: Big data
  text-oriented benchmark creation for {Hadoop}. {IBM} Journal of Research and
  Development  57(3/4),  10:1--10:6 (2013)

\bibitem{Gray1993}
Gray, J.: The Benchmark Handbook for Database and Transaction Systems (2nd
  Edition). Morgan Kaufmann (1993)

\bibitem{Guille2015}
Guille, A., Favre, C.: Event detection, tracking, and visualization in twitter:
  a mention-anomaly-based approach. Social Network Analysis and Mining  5(1),
  ~18 (2015)

\bibitem{Kilinc2017}
Kılın{\c{c}}, D., {\"{O}}z{\c{c}}ift, A., Bozyigit, F., Yildirim, P.,
  Y{\"{u}}calar, F., Borandag, E.: {TTC-3600:} {A} new benchmark dataset for
  turkish text categorization. Journal of Information Science  43(2),  174--185
  (2017)

\bibitem{Lewis2004}
Lewis, D.D., Yang, Y., Rose, T.G., Li, F.: {RCV1:} {A} new benchmark collection
  for text categorization research. Journal of Machine Learning Research  5,
  361--397 (2004)

\bibitem{Manning2008}
Manning, C.D., Raghavan, P., Sch{\"u}tze, H.: Introduction to information
  retrieval. Cambridge University Press (2008)

\bibitem{OShea2010}
O'Shea, J., Bandar, Z., Crockett, K.A., McLean, D.: Benchmarking short text
  semantic similarity. International Journal of Intelligent Information and
  Database Systems  4(2),  103--120 (2010)

\bibitem{Paltoglou2010}
Paltoglou, G., Thelwall, M.: A study of information retrieval weighting schemes
  for sentiment analysis. In: 48th Annual Meeting of the Association for
  Computational Linguistics. pp. 1386--1395 (2010)

\bibitem{Partalas2015}
Partalas, I., Kosmopoulos, A., Baskiotis, N., Arti{\`{e}}res, T., Paliouras,
  G., Gaussier, {\'{E}}., Androutsopoulos, I., Amini, M., Gallinari, P.:
  {LSHTC:} {A} benchmark for large-scale text classification. CoRR
  abs/1503.08581 (2015)

\bibitem{Ravat2008}
Ravat, F., Teste, O., Tournier, R., Zurfluh, G.: Top\_keyword: an aggregation
  function for textual document {OLAP}. In: 10th International Conference on
  Data Warehousing and Knowledge Discovery (DaWaK). pp. 55--64 (2008)

\bibitem{Reagan2015}
Reagan, A.J., Tivnan, B.F., Williams, J.R., Danforth, C.M., Dodds, P.S.:
  Benchmarking sentiment analysis methods for large-scale texts: {A} case for
  using continuum-scored words and word shift graphs. CoRR  abs/1512.00531
  (2015)

\bibitem{SparckJones2000a}
{Spärck Jones}, K., Walker, S., Robertson, S.E.: A probabilistic model of
  information retrieval: development and comparative experiments: Part 1.
  Information Processing \& Management  36(6),  779 -- 808 (2000)

\bibitem{SparckJones2000b}
{Spärck Jones}, K., Walker, S., Robertson, S.E.: A probabilistic model of
  information retrieval: development and comparative experiments: Part 2.
  Information Processing \& Management  36(6),  809 -- 840 (2000)

\bibitem{Truica2016}
Truică, C.O., Darmont, J., Velcin, J.: A scalable document-based architecture
  for text analysis. In: International Conference on Advanced Data Mining and
  Applications (ADMA). pp. 481--494 (2016)

\bibitem{Wang2016}
Wang, L., Dong, X., Zhang, X., Wang, Y., Ju, T., Feng, G.: {TextGen:} a
  realistic text data content generation method for modern storage system
  benchmarks. Frontiers of Information Technology \& Electronic Engineering
  17(10),  982--993 (2016)

\end{thebibliography}

\end{document}